# Exploring diversity perceptions in a community through a Q&A chatbot

Peter Kun[a,*], Amalia De Götzen[a], Miriam Bidoglia[b], Niels Jørgen Gommesen[a], George Gaskell[b]

[a]Aalborg University Copenhagen, Denmark
[b]London School of Economics and Political Science, UK

*corresponding e-mail: peter@peterkun.com

**Abstract:** While diversity has become a debated issue in design, very little research exists on positive use-cases for diversity beyond scholarly criticism. The current work addresses this gap through the case of a diversity-aware chatbot, exploring what benefits a diversity-aware chatbot could bring to people and how do people interpret diversity when being presented with it. In this paper, we motivate a Q&A chatbot as a technology probe and deploy it in two student communities within a study. During the study, we collected contextual data on people's expectations and perceptions when presented with diversity during the study. Our key findings show that people seek out others with shared niche interests, or their search is driven by exploration and inspiration when presented with diversity. Although interacting with chatbots is limited, participants found the engagement novel and interesting to motivate future research.

## 1. Introduction

Diversity has become a widely debated topic in design, particularly when referring to new digital technologies, representation, and democracy (Costanza-Chock, 2020). The European Commission's Ethics Guidelines for Trustworthy AI (2019), among others, emphasizes that AI should support diversity and the subjective well-being of people. However, while scholars and ethicists highlight issues around diversity (e.g., Keyes, 2019; Matzner, 2019; Schelenz et al., 2019), best practices for positive use-cases of diversity remain less discovered. In this paper, we address this lack of knowledge by focusing on what people expect when presented with diversity. Our approach is to study people's diversity perceptions and expectations when faced with a technological artifact, positioning our work in the Human-Computer Interaction (HCI) field. To study this phenomenon with a research-through-design

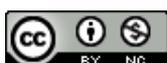

*Peter Kun, Amalia De Götzen, Miriam Bidoglia, Niels Jørgen Gommesen, George Gaskell*

approach (Zimmerman et al., 2007; Stappers and Giaccardi, 2017), we propose and develop a technology probe, a Question-and-Answer chatbot, with a diversity focus and deploy it in two student communities. The leading research question for the study is the following:

> **RQ1:** What do people expect from being presented with diversity? How do people interpret diversity in use?

We also generate contextual knowledge during the research-through-design procedure about our users' usage of our chatbot technology probe. We formulate this as the following questions:

> **RQ2:** How do people perceive a chatbot that connects community members through diversity attributes?

> **RQ3**: What do student communities ask from others through a chatbot?

Next, we sharpen our frame on diversity.

## 1.1 Framing diversity

Diversity exists between individuals and emerges through interaction: we can recognize and qualify diversity by comparing ourselves to others. Despite being a seemingly easy concept to grasp, diversity is a complex, multi-layered compositional construct. Diversity comes from the differences in attributes between individuals, and diversity also goes beyond physical or behavioral attributes (Jackson et al., 1995). Thus, diversity goes beyond stereotypes, and scholars have differentiated between surface-level or deep diversity, or in other words, their demographical and attitudinal diversity (Harrison et al., 1998). Today's technology can observe, register, and learn those behaviors through the many data traces we leave behind ourselves, but instead of leveraging that diversity, it is actively reducing it. Filter bubbles are a known limitation of social media, where like-minded people get to interact with each other and with the same information (Pariser, 2012). It seems diversity issues are fostered and increasingly surfaced through how we design technology to connect people locally or globally.

**Diversity as a central element in technology design**

Despite the increasing focus on diversity in Human-Computer Interaction (HCI) in recent years, researchers tend to consider only a limited number of diversity dimensions (Himmelsbach et al., 2019, p.12). Himmelsbach et al. (ibid, p.10) propose considering diversity dimensions from the users' lived experience and identity, the context of use, and everyday experiences at the center of human-centered design. To create a more differentiated understanding of user diversity, they propose that future studies focus on analyzing relations between diversity dimensions, including the dimensions predominantly addressed together (ibid, 2019, p.12). Dankwa and Draude's framework (2021) centers on diversity at the core of human-centered design, leaving behind practices and methodologies from the past's colonialist ideals within HCI, including the structural and institutionalized forms of oppression and biases inherent in the design of technology. They conclude that successful adoption of the framework demands: 1) confronting and redesigning the embedded structures and repressive institutions that enforce inequality; 2) questioning the field's methodological approaches, practices, and systems of knowledge; and 3) advocating for the empowerment of the users.

Bardzell (2010) arrives at a similar position by showing that feministic theories and methods can encourage different sensibilities to design for cultural differences and constructive engagement with diversity in all HCI research and design process stages. Fletcher-Watson et al. (2018) claim that adopting a moral and ethical position of accepting difference in diversity computing can facilitate mutual understandings, enhance inclusion, and guide social interactions (2018, p.32). One way forward for creating diversity computing that challenges human cognitive biases, they argue, is to let people participate in other's sense-making activities through participatory methods, self-reflection, and reflection with others (Fletcher-Watson et al., 2018, p.32). Interestingly, Fletcher-Watson et al.'s (2018) diversity computing vision argues against the technological issues that large-scale data processing introduces, such as bias. Moreover, it provides an agenda to embrace differences between people instead of eliminating them. As Fletcher-Watson et al. (2018) propose, the more constructive framing of diversity in computing is to reframe human-computer interaction to human-human technologically-mediated interaction to incorporate diversity principles into computing. Next, we consider the understanding of diversity in technologically-mediated contexts.

## 1.2 Technologically-mediated diversity

Computer scientists have taken further the natural sciences' initial formal description of species diversity (Simpson, 1949) for recommendation systems (e.g., webshops and music streaming services). When a user searches for a product to buy or a song to listen to, the recommendation system attempts to show an accurate result (Pazzani and Billsus, 2007). However, recommendation systems have also considered objectives beyond accuracy for ambiguous queries, such as diversity, serendipity, and novelty (Kaminskas and Bridge, 2016). For an ambiguous query, diverse results mean a wider spread of results to increase the chances of success. Serendipity and novelty objectives focus on surprising the user with a relevant recommendation that the user did not expect, which may also be unknown. Although these objectives' mathematical and algorithmic formulation is possible, they are still context-dependent and require experiments to evaluate their utility within a given context (Kaminskas and Bridge, 2016). However, it remains unclear how to operationalize diversity in the specific context of studying users' perceptions when presented with diversity.

To conclude, the current state-of-the-art of diversity in the fields of design and HCI highlights issues of earlier technology-first diversity solutions and provides scholarly criticism without elaboration on positive use-cases for diversity, that designers could apply as best practices. Meanwhile, diversity in the context of recommendation systems provide an expanded view of "beyond accuracy" measures, including diversity, serendipity, or novelty in recommendations. Yet, this literature lacks use-cases of connecting people through diversity attributes and supporting communities with social recommendations beyond accuracy.

## 2. Developing a Q&A Chatbot

This section elaborates on our design rationale for a Q&A chatbot that we developed as a technology probe for the research-through-design process to investigate users' perceptions and expectations of being presented with diversity.

### *2.1 Chatbots*

Early research on chatbots, such as ELIZA (Weizenbaum, 1966) or Alice (Wallace, 2009), focused on having an artificial intelligence system interact with a human in natural language. While earlier chatbots could not deliver on such promises, current integrated chatbot systems on social media networks and instant messaging platforms like Facebook Messenger or Slack have induced new optimism. Seering et al. (2019) provide a typology of the last wave of chatbots to investigate them from a community perspective, depending on a chatbot's engagement type. They characterize chatbots designed for:

1. Dyadic chatbots (a chatbot having a one-on-one conversation);
2. Broadcasting chatbots, chatbots that send messages to many users simultaneously but otherwise do not engage in a conversation; and
3. Multiparty-based chatbots engaged in back-and-forth conversations involving multiple users, like participants in a group conversation.

Opposed to intelligent agents and conversation partner chatbots, Klopfenstein et al. (2017) described *"botplications"* as a new generation of chatbots that are small thread-based interfaces to fulfill simple functions that would not necessarily warrant a separate app. These botplications contrast historical counterparts of ELIZA and ALICE but are valuable software solutions to help the user solve specific, narrow tasks.

### *2.2 Q&A*

Community Q&A sites, such as Yahoo!, Answers or StackOverflow, have been extensively studied in Human-Computer Interaction on their function in sharing expertise and knowledge management (Ackerman et al., 2013) or using informational and conversational needs of users (Harper et al., 2009). The latter category of social Q&A sites has declined in popularity in favor of other platforms such as Facebook or Twitter while fulfilling social Q&A needs through public posts to friends (Morris et al., 2010). To conclude, the essential concept of Question Asking and Answering has been continuously transitioning from older to newer social platforms on the internet, making the core Q&A concept still relevant today.

## 2.3 Designing a Q&A chatbot: Technology probe

To study user perceptions in a community when presented with diversity, we developed a technology probe (Hutchinson et al., 2003), a Q&A chatbot named the AskForHelp chatbot. Technology probes enable studying users' needs in a real-world setting while enabling field-testing of technology and inspiring people to reflect on new technologies (Hutchinson et al., 2003). We approached the AskForHelp chatbot from the vein of a *"botplication"* (Klopfenstein et al., 2017), a lightweight interface that connects people in a community to avoid the technical complexity of a full-fledged conversational agent with natural language processing and training data. Our investigation of the state-of-the-art of what would be appropriate algorithms in the context of studying users' perceptions and expectations from being presented to diversity showed limited existing knowledge – especially when it concerns such "beyond-accuracy" objectives. Therefore, the current study does not implement an algorithm but approaches it like a Wizard-of-Oz intervention (e.g., Nordberg et al., 2020). However, instead of a researcher acting like a chatbot, we mean that the chatbot's *matchmaking logic is randomized*. Due to our research aim of collecting data with the chatbot, we chose the dyadic interaction model (Seering et al., 2019), where the chatbot is the single interface for all users, facilitating any user-user interaction.

Unlike other studies on Q&A using existing social media platforms to study Q&A behavior (e.g., Harper et al., 2009; Morris et al., 2010), we have approached our research questions with a technology probe (Hutchinson et al., 2003) specifically designed and developed for our research questions. Such approach enabled us to approach the design of our technology probe based on the following design principles:

- **Focus on one core interaction – question-asking and answer-giving:** We have prioritized design decisions to serve the core interaction between two individuals in a community, without other common social features. Delimiting user-user interactions emphasize the research instrument characteristics of the chatbot, focusing on our research questions, and decreasing confounding variables.
- **Motivate engagement:** We have introduced a gamification mechanism in the chatbot that sends out nudging messages if a user does not interact for an extended period. We chose this approach to motivate engagement for question-asking and answer giving for ensuring sufficient number of interactions with the technology probe.
- **Emphasize diversity:** The tone of voice and our communication about the chatbot and study have aimed to emphasize to take mutual benefit from the diversity in a community.

*Peter Kun, Amalia De Götzen, Miriam Bidoglia, Niels Jørgen Gommesen, George Gaskell*

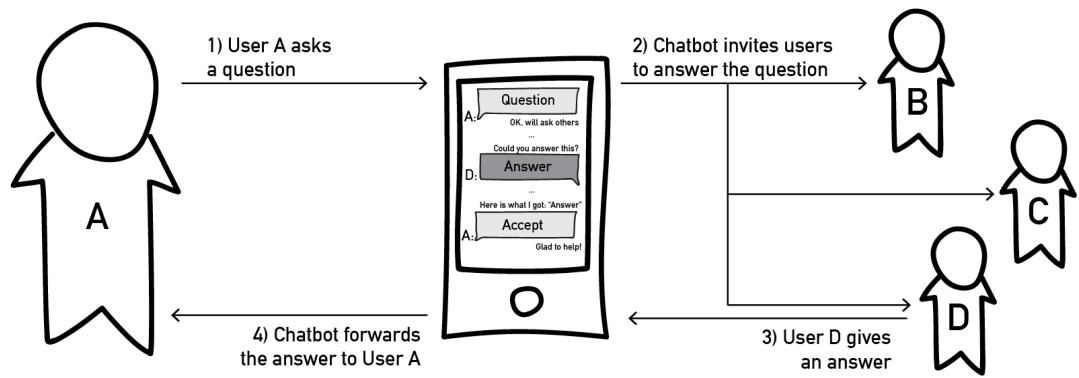

*Figure 1. The user-user interaction is entirely mediated through the chatbot.*

We operationalized these design principles as the following:

**User-user interaction model**

Following these design principles, the user-user interaction model we designed is shown in Figure 1. When User A wants to ask a question, she triggers the /question command in the first step. The chatbot invites other community members (Users B, C, and D) to answer the question in the second step. Any and multiple of these users can answer. In the third step, User D answers. In the fourth step, the chatbot forwards User D's answer to User A, who can then accept the answer or ask the chatbot to invite more users to answer.

**Delimited user-user interactions**

The users only see each other's first names and have no other direct contact. The first names can be filled with fictional or pseudonymous names, enabling anonymous interaction on-demand. There is no way of following up on a question or seeing the answers of others, making the interaction limited and one-shot.

**Diversity prompts**

We have introduced "diversity prompts", chatbot messages sent with potentially interesting diversity facts about the student populations from an earlier study. We expected these prompts to trigger participants to ask questions.

**Random algorithm**

We chose to use Wizard-of-Oz on the chatbot's algorithm side, meaning no specific recommendation algorithm logic was in place; a Questioner user was randomly connected to Answerer users. This methodological choice enabled us to use the chatbot as a research instrument primarily without the confounding factor of a matchmaking algorithm between users. This choice also supported our intention to use the study's findings to design and develop such an algorithm in the future.

## 3. Method

We conducted a study on two sites to investigate what people expect when presented with diversity through a technology probe. In the following, we outline the methodology of this study.

*3.1 Participants*

We deployed the chatbot in two communities simultaneously in two different universities; Aalborg University Copenhagen (Pilot A) and London School of Economics (Pilot B). Pilot A ran with 34 participants, and Pilot B ran with 46 participants; overall, 80 university students participated in our study. Both pilots ran for two weeks, and we financially compensated for the active involvement of participants. In Pilot A, participating students received the equivalent of about 20 EUR for their participation. In Pilot B, a similar amount of money was offered to charity on behalf of the participants.

*3.2 Setup*

In the beginning, we informed all participants about the study's objective and duration, the ways of data collection, and how personal information and data are handled. Participating students could contact the researchers behind the project via a designated email address.

The participants needed to install Telegram on their phones and install the AskForHelp chatbot, as shown in Figure 2.

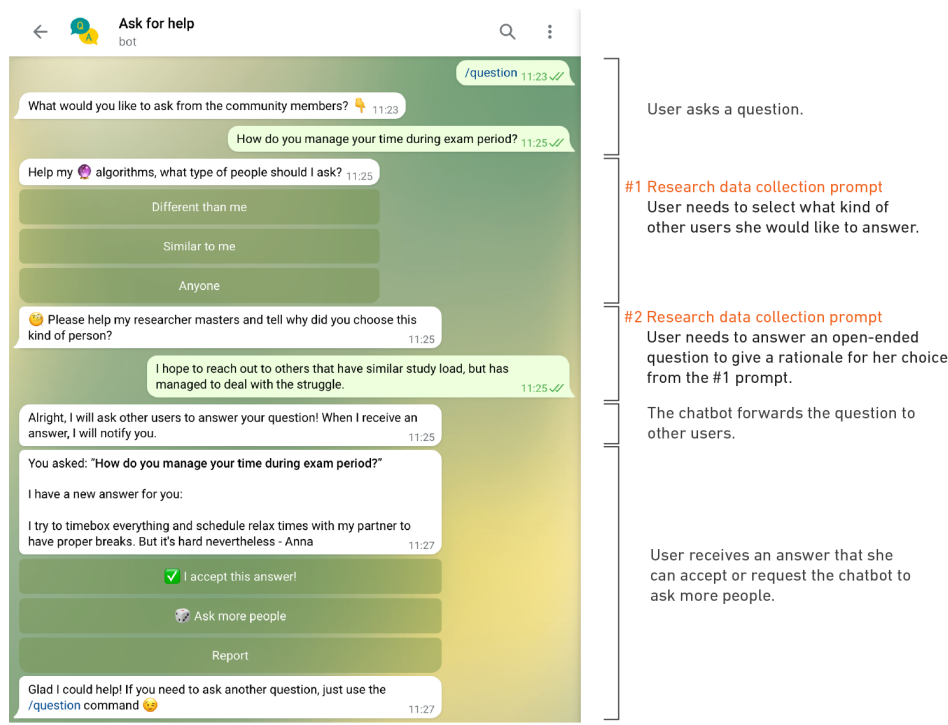

*Figure 2. The chatbot mediated the interaction between users. Within the conversation, we also placed research prompts (#1 and #2 highlighted) for data collection to address our research questions.*

*Peter Kun, Amalia De Götzen, Miriam Bidoglia, Niels Jørgen Gommesen, George Gaskell*

*3.3 Data collection and analysis*

We approached the data collection with mixed methods, which we will provide an overview of below.

**Thematic analysis from log files**

We collected 1) log files, 2) a database of the participants' questions and answers, and 3) timestamped data on all interactions throughout the chatbot. These were thematically analyzed.

**In-conversation research questions**

We injected data collection questions into the conversation flow whenever a user asked a question; see Figure 2 for the two research prompt questions. The conversation:

| | |
|---|---|
| [Chatbot]: | What would you like to ask from other users? |
| [User]: | [user enters a question] |
| [Chatbot]: | Help my algorithms, what type of people should I ask? |
| [User]: | User chooses one from three buttons: [Similar][Different][Anyone] |
| [Chatbot]: | Why did you choose this answer? |
| [User]: | [user provides an answer on diversity expectation] |

The choices between similar/anyone/different and the answers provided qualitative data within context. We hypothesized that these questions would provide contextual answers to understand what people expect when presented with diversity.

**Exit survey and UTAUT2 questionnaire**

We held a survey (85% fill rate, 68 out of 80 participants) at the end of the study. The survey contained a tailored UTAUT2 questionnaire to assess the participants' views on the chatbot design and its integration into their everyday life. The UTAUT2 (Unified Theory of Acceptance and Use of Technology) questionnaire is a comprehensive instrument for technology acceptance (Venkatesh et al., 2012), measuring performance, effort, and enjoyment, among other factors. We used UTAUT2 for measuring to what degree the AskForHelp chatbot as a technology probe was a successful instrument used for our research inquiry and to understand its influence on our findings.

**Focus groups - Understanding the participants' chatbot experiences**

We held 2+2 focus group interviews on the two pilot sites, with 4-5 study participants participating in each interview. The four focus group sessions were recorded, transcribed verbatim, and analyzed with the other collected data for triangulation.

The focus group sessions aimed to 1) get feedback on the participants' experiences with the chatbot, 2) capture individual concerns about chatbot use, and 3) to provide participants with the opportunity to elaborate on their personal experiences with the chatbot and how it influenced their participation and communication with others in the study.

## 4. Results

Over the two weeks of engagement with the chatbot, Pilot A participants asked 402 questions and sent 1638 answers, while Pilot B participants asked 257 questions and sent 762 answers, overall 669 questions and 2400 answers in the whole study. To address *RQ3 "What do student communities ask from others through a chatbot?"*, we conducted a systematic content analysis on the asked questions to classify them based on their apparent communication goal or request aim. Finally, we developed a coding scheme iteratively to categorize the questions into mutually exclusive types. Table 1 shows the seven identified categories with examples and the distribution of question types.

*Table 1.    Questions asked through the chatbot varied over topics.*

| Question type | Example question | Pilot A | Pilot B |
| --- | --- | --- | --- |
| Information | "Is the [park] open?" | 4% | 4% |
| Community | "Are you thinking of staying in [city] after graduating? | 54% | 32% |
| Connection | "Anyone interested in forming a study group?" | 0% | 2% |
| Opinions and experiences | What do you think about the vaccine delivery condition in [country]? | 14% | 15% |
| Suggestion | "Any fiction book recommendations?" | 22% | 30% |
| Academic | "Any tips for summer exams?" | 4% | 14% |
| Personal or sensitive | "Do you want to get married in the future? Do you want kids?" | 1% | 3% |
| N | | 401 | 246 |

### 4.1 Assessing the chatbot

To ensure the validity of our results with the chatbot and to address RQ2: *"How do people perceive a chatbot that connects community members through diversity attributes?"*, we first assess with the UTAUT2 instrument (Venkatesh et al., 2012) to what extent the AskForHelp chatbot was found appropriate and usable. Table 2 summarizes the exit survey results.

*Peter Kun, Amalia De Götzen, Miriam Bidoglia, Niels Jørgen Gommesen, George Gaskell*

*Table 2. Descriptive evaluation of the chatbot user experience. Darker green represents a higher mean.*

| Question/means (between 1..5, 5 is highest) | Pilot A | Pilot B |
|---|---|---|
| A) It was easy to install the chatbot | 4 | 4.1 |
| B) It was easy to ask a question in the chatbot | 4.1 | 4.1 |
| C) It was easy to provide an answer in the chatbot | 4.1 | 4.1 |
| D) It was easy to decide if I liked an answer | 3.8 | 3.6 |
| E) I had the necessary resources to use the chatbot | 4.1 | 4.2 |
| F) I had the necessary knowledge to use the chatbot | 4.3 | 4.1 |
| G) Chatbot helped me to acquire new ideas | 3.2 | 3.7 |
| H) Chatbot was useful to reach out for help | 3.6 | 4 |
| I) Chatbot was useful to provide help to others | 3.8 | 4.1 |
| J) Chatbot useful to get to know other students | 3.1 | 2.7 |
| K) Chatbot useful to make me feel part of a community | 3.4 | 3.2 |
| L) I felt comfortable using the chatbot to ask questions | 3.9 | 4.1 |
| M) I felt comfortable using the chatbot to answer questions | 4.2 | 4.2 |
| N) I felt pleased to be able to provide an answer | 4.4 | 4.2 |
| O) I felt pleased to get answers to my questions | 4.2 | 4.1 |
| P) Chatbot had an appealing tone of voice | 3.4 | 3.5 |
| Q) I found the chatbot trustworthy | 3.7 | 3.6 |
| R) Using the chatbot was rewarding | 3.3 | 3.6 |
| S) Using chatbot was fun | 3.6 | 3.6 |
| T) I was interested in the experience of chatbot | 4.1 | 4.1 |
| U) I would keep using the chatbot in my everyday life | 2.8 | 3.2 |
| V) I use other chatbots in my everyday life | 1.8 | 1.8 |

In the UTAUT2 instrument, questions A to D are related to effort expectancy, questions E and F to facilitating conditions, questions G to K to performance expectancy, questions L to T to hedonic motivations, and question U investigates behavioral intentions, and the final question V is related to habits. We discovered that, in general, our participants found the chatbot intuitive to use, a meaningful way to connect to a community, and an enjoyable experience. We discuss these below together with insights gained from the focus group interviews.

**User-user interaction model**

Our findings show that participants found the chatbot easy to use (see A to F), indicating that a chatbot approach is suitable for student communities, even when users otherwise do not use chatbots in everyday life (see V). The limited interaction model was found controversial during the focus groups. Some participants preferred to follow up with someone who answered their questions, engaged with the same person over many questions, or even wished to exchange contact info and take their conversation out of the chatbot. Participants described this as the chatbot being annoying and interruptive for "real conversations". However, this perception declined as they got used to the interaction model. One participant found the constrained interaction model "refreshing" because it steered her to answer questions more often without the need to engage in a larger conversation. The participants found, in general, that the chatbot facilitated a novel interaction with fellow students, especially first-year students who started their education in remote classrooms due to COVID-19 lockdowns.

**Notification amounts and sensitive topics**

Participants unanimously mentioned the issue of simply receiving too many notifications from the chatbot, which they often solved by muting it for parts of the day. Another user highlighted how she had difficulty giving private answers to sensitive questions. While it is intriguing not to talk directly to a person through the chatbot, she caught herself stopping a reply realizing that she usually would not share such personal thoughts with a stranger.

**Connecting individuals to the community**

Statements J and K show a moderately positive sentiment on the chatbot as a valuable instrument to know more or feel part of the community, while statements H and I show the chatbot's usefulness in reaching out for help.

**Enjoying the experience**

Statement L to S explores the more hedonic motivation in using the chatbot. Most participants felt at ease and enjoyed asking and replying to questions while finding the chatbot experience engaging. Nevertheless, one critical aspect is related to the tone of voice used in the chatbot and explored in statement P, which we believe is also related to the moderately positive judgment concerning its trustworthiness (statement Q).

The results of the UTAUT2 instrument show that the chatbot fulfilled its role as a technology probe. The participants found it a reasonable interface to connect with others in their community. Furthermore, the UTAUT2 instrument shows a good user experience, providing confidence that usability issues and similar distractions do not clutter the collected data through the chatbot. However, as indicated in statement U), the participants would not use the chatbot in its current form in their everyday lives.

The following section highlights our findings on the users' expectations and perceptions of diversity.

*Peter Kun, Amalia De Götzen, Miriam Bidoglia, Niels Jørgen Gommesen, George Gaskell*

*4.2 Expectations and perceptions of diversity*

To address RQ1: *"What do people expect from being presented with diversity? How do people interpret diversity in use?"*, within each chatbot conversation, we asked the study participants to provide a rationale for what kind of user they would like them to answer, from the choices: "Similar to me", "Different than me", "Anyone". Following these research prompt questions, we asked the users open-ended answers on why they preferred this and what they expected from this selection (see Figure 2). Table 3 presents the distribution of choices made. As "Anyone" was a possible choice, we were not surprised that the dominant number of questions (73%) selected this. Instead, we primarily focused on the cases when users made up their minds between "similar" or "different" and gave an eloquent answer, not just a quick trivial answer.

Table 3.   Distribution of choices made by the participants on the #1 research prompt.

| "What type of people should I ask?" | | |
|---|---|---|
| Ask… | Pilot A | Pilot B |
| Anyone | 72% | 74% |
| Different to me | 8% | 11% |
| Similar to me | 20% | 14% |
| N | 402 | 257 |

We conducted a content analysis based on answers from the #2 research prompt while conversing with the chatbot (see Figure 2), categorizing them into eleven mutually exclusive categories based on their diversity expectations. We pooled the data from Pilot A and Pilot B for this analysis. The identified categories were:

- **Taste (similar/different):** Finding people with similar or different tastes. Taste dominantly referred to categories where people asked for personal recommendations, such as music or a TV show recommendation.
- **Life experience (similar/different):** Finding people with similar or different life experiences. Under life experience, we categorized questions such as someone looking for different people's experiences in making money online or looking for others' experiences that have lost their jobs due to the COVID-19 pandemic.
- **How are other humans (similar/different):** In this category, we identified more contemplative questions, probing similar people about what they think about specific issues or looking for answers from people who have different opinions or world-view on different topics.

- **Curiosity (similar/different/anyone):** Participants answered curiosity very often, and we found it to be a choice to avoid giving an elaborative, more thoughtful answer.
- **Different ideas for concrete needs (different):** covering exact needs, like quickly getting a gluten-free cake.
- **N/A (similar/different/anyone):** Answers in this category were the users opting out from giving a proper answer, such as typing in a single dot ".".
- **No filter (anyone):** Answers in this category contained general comments, such as "my question is so general, I did not want to specify it further".
- **Diverse + more (anyone)**: Answers in this category focused on having as diverse answers as possible.
- **Meta-questions about chatbot (anyone):** These answers referred to the workings of the chatbot.
- **Similar study reasons (similar):** Answers in this category were expected to be relevant to people from the same studies.
- **Similar in self-identity (similar):** Answers in this category focused on finding other users with similar niche interests.

Table 4 shows the eleven identified categories with examples and the distribution of question types.

*Table 4. Counting the coded diversity expectation answers of the #2 research prompt.*

| Category | Example | Similar | Different | Anyone | N |
|---|---|---|---|---|---|
| Taste | "Because they have the same taste hopefully" (similar) | 22 | 4 | - | 26 |
| Life experience | "Wanted to know if they had similar jobs like me" (similar) | 5 | 5 | - | 10 |
| How are other humans? | "Nice to get opinions that may differ from your own" (different) | 14 | 17 | - | 31 |
| Curiosity | "For curiosity" (different) | 14 | 11 | 26 | 51 |
| Concrete needs | "Need a local" (different) | - | 10 | - | 10 |
| Meta - chatbot | "Not sure what it means, not sure how you have profiled me" (anyone) | - | - | 35 | 35 |
| Study reasons | "Would like to hear the opinion of other master students at [uni]" (similar) | 28 | - | - | 28 |
| Self-identity | "I want to ask others who are into video games" (similar) | 12 | - | - | 12 |
| No filtering | "Want to get as many answers as possible" (anyone) | - | - | 218 | 218 |
| Diverse + more | "I want to hear different points of view" (anyone) | - | - | 70 | 70 |
| N/A | "Masters" (anyone) | 18 | 12 | 126 | 156 |
| N | | 113 | 59 | 475 | |

## 5. Discussion

This section interprets our findings, characterizes design implications, highlights limitations, and discusses potential future work.

*5.1 Design implications of people's diversity perceptions*

**Similarity**

The participants unanimously found it hard to interpret the seemingly straightforward yet paradoxical question: *"Who are people similar to me?"* In our analysis, the most constructive frames on similarity were considering those that share niche interests with them or see the world through similar values. Clear examples of this interpretation were answers where people inquired from other people with deep interest about a specific phenomenon, such as veganism, subcultures, or specific video games. In other words, the participants expected to connect with people diverse in personal, deep-level diversity attributes (Harrison et al., 1998). Members can share a specific niche interest beyond similarity or difference in surface-level diversity in a community.

The design implication of catering to this human need to reach out to others that share niche interests means that the system needs fine-grained diversity characterization of the users and lets people reach out to "their tribes", even when those are implicit.

**Difference**

Key findings from our analysis show that people had fewer issues characterizing and interpreting who are "different" from them. Despite our expectation that people will seek the opinions of significantly different people than them, we found few occasions of such expectation. Most of the expectations carried an exploratory quality, where people were driven by curiosity or wished for serendipity in asking people different from themselves. Another interpretation of difference came from a "knowledge gap" perspective; a person asking a question did not know the answer and interpreted "different" as seeking someone who knows.

The design implication to cater to an exploratory need of difference can be fulfilled with beyond-accuracy algorithm goals, such as diversity, serendipity, or novelty in query results. However, there is also an opportunity for future research to better understanding the motivations and typology of *what kind* of different people would aspire to reach.

**Profiling and algorithms**

Diversity-aware algorithms need to consider some user profiling to compare other users to the user (Furtado et al., 2013). This profiling should cover meaningful attributes users identify with and compare themselves with others. Furthermore, different types of diversity expectations could be computationally modeled, such as a user seeking inspiration or challenging views, which can be a base for providing user control over a diversity-aware algorithm.

Our study illustrates one approach to incorporate diversity principles into computing as proposed by Fletcher-Watson et al.'s (2018), to frame diversity in computing as human-human technologically-mediated interaction. Overall, our study shows that social Q&A is a valuable context to study beyond-accuracy recommendation system goals, that involve diversity, serendipity, and novelty (Kaminskas and Bridge, 2016). When people ask questions, recommendations, or opinions from each other, there can be multiple specific answers, thus beyond-accuracy becomes more important than when the user seeks ground truth as an answer. Such algorithms need to be based on user profiling of users (Furtado et al., 2013) to have diversity attributes available as a design material. While the user profiling is a necessary step to provide data for algorithms, and it is by nature a normative approach with potential caveats of biasing (Costanza-Chock, 2020; Dankwa and Draude, 2021), our study provides a more nuanced understanding of what people consider diverse. Future studies may build on our findings for meaningful user profiling, as opposed to the normative, colonizing, and biasing ones. Such diversity profiling can be based on rich diversity dimensions based on users' lived experience and identity and everyday experiences, which highlight that people are diverse in unlimited ways, and strive for a decolonizing such algorithms, by featuring attributes that go beyond surface-level attributes, across cultures, backgrounds, upbringing, and so forth.

## 5.3 Limitations

Key findings show that a Q&A chatbot can present people with diversity; however, the reported study has its limitations: Firstly, the pilots ran as paid research experiments, and both pilots ran during COVID-19 lockdown periods. We gathered limited information on how people would find long-term to be presented to diversity through a chatbot when they can also meet with others physically. Furthermore, the chatbot's limitations on user-user interactions are uncommon in other social network sites, and the research prompts to capture our research data were tedious to answer. Last, while we promoted and communicated the chatbot to connect two users based on diversity dimensions, the current study featured a random algorithm. With this algorithm, we could not evaluate the quality of connections established.

*Peter Kun, Amalia De Götzen, Miriam Bidoglia, Niels Jørgen Gommesen, George Gaskell*

## 6. Conclusions

When people are presented with diversity, we can conclude that they seek similarity and difference in deep-level diversity attributes. We studied this phenomenon through a chatbot, a technology probe, deployed in two student communities. When people are presented with diversity, they seek others with similar niche interests or seek different views of different others for curiosity. Our study indicates that there can be positive use-cases for diversity.

In the future, the next iteration of the chatbot will implement diversity-aware algorithms based on our current explorations. Such diversity-aware algorithms will allow the participants to specify what kind of diversity they would like to experience with the chatbot, moving from the current random algorithm. With this future development, the chatbot will enable a more detailed study of how people would like to be presented with diversity when they can "steer" the algorithm. Additionally, from a methodological point of view, we will also deploy the chatbot in students communities in multiple countries to gain a more grounded perspective on how a diversity-centered chatbot is perceived in different cultures.

**Acknowledgments:** The presented research is part of the WeNet project, which is funded by the European Commission under the Horizon2020 research funding scheme "Future and Emerging Technologies (FET)", under the grant agreement 823783.
The authors would like to express their appreciations for the members of the WeNet consortium for the fruitful collaboration leading up to the presented work, and to the students participating in the study.

*Peter Kun, Amalia De Götzen, Miriam Bidoglia, Niels Jørgen Gommesen, George Gaskell*

About the Authors:

**Peter Kun** is Postdoctoral Researcher and member of the Service Design Lab at Aalborg University. In the WeNet project, Peter researches human-AI interactions for diversity. His broader research agenda combines data science techniques with design for creative uses of data.

**Amalia De Götzen** is Associate Professor at Aalborg University in Copenhagen and a member of the Service Design Lab. Amalia's research activity focuses on Digital Social Innovation and in particular on the intersection between Interaction Design and Service Design.

**Miriam Bidoglia** is Research Assistant at London School of Economics and Political Science. In the WeNet project, Miriam supports the running of the pilots in London, and contributes to the research with quantitative and qualitative data analysis.

**Niels Jørgen Gommesen** is Research Assistant in the WeNet project at Aalborg University. He is PhD-student in Media Science, focusing on communication in citizen science, and designing more responsible forms of citizen engagement strengthening cultural and scientific citizenship.

**George Gaskell** is Professor Emeritus in Social Psychology and Research Methodology at London School of Economics and Political Science. In the WeNet project, George contributes to the sociological research agenda around diversity, survey design, and quantitative and qualitative research design.